\documentstyle[11pt,aaspp4]{article}

\newcommand{\X}{{\rm X}}
\newcommand{\Y}{{\rm Y}}

\newcommand{\Zn}{{\rm Zn}}
\newcommand{\Cr}{{\rm Cr}}
\newcommand{\Fe}{{\rm Fe}}
\newcommand{\Ni}{{\rm Ni}}
\newcommand{\Si}{{\rm Si}}
\newcommand{\Xg}{{\rm X,g}}

\newcommand{\Xd}{{\rm X,d}}
 
\newcommand{\HI}{{{\rm H}\,{\rm I}}}
\newcommand{\Hy}{{\rm H}}
\newcommand{\ism}{{\rm ism}}
\newcommand{\Gal}{{\rm Gal}}
\newcommand{\dla}{{\rm dla}}

\newcommand{\obs}{{\rm obs}}
\newcommand{\intr}{{\rm dla}}

\newcommand{\ctilde}{\tilde{c}}
\newcommand{\Ztilde}{\tilde{Z}}
\newcommand{\nutilde}{\tilde{\nu}}
\newcommand{\ktilde}{\tilde{k}}

\lefthead{Giovanni Vladilo}
\righthead{Dust and elemental abundances in Damped Ly$\alpha$ absorbers}

\begin{document}

\title{Dust and
elemental abundances in Damped Ly$\alpha$ absorbers}

\author{Giovanni Vladilo\altaffilmark{1} }
\affil{Osservatorio Astronomico di Trieste, Via G.B. Tiepolo 11,
        I-34131,  Trieste, Italy }

\begin{abstract}

The effects of the dust on the
determination of elemental abundances in damped Ly $\alpha$
(DLA) absorbers are investigated.  
Relations between
the observed abundances measured in the gas phase 
and the overall abundances (gas plus dust) 
are derived as a function of 
dust-to-gas ratio, metallicity, element-to-element abundance
pattern, average extinction coefficient of dust grains,
and chemical composition of dust grains. 
A method is presented for determining
dust-to-gas ratios, dust-to-metals ratios, and
dust-corrected relative abundances in DLA absorbers 
by assuming dust of Galactic type
and constant abundance ratios between iron-peak elements.
The method is applied to a sample of
17 DLA absorbers  with available   Zn, Cr and/or Fe
measurements. 
The resulting  dust-to-gas ratios are mostly distributed
between 2\% and 25\% of the Galactic value, 
in good quantitative agreement
with the results from  reddening  studies of 
QSOs with foreground DLA absorption. 
A correlation is found between dust-to-gas ratio and
metallicity in DLA galaxies, with a typical
dust-to-metals ratio of $\approx$ 60\% the Galactic value.  
The derived dust-to-metals ratios  
are then used to correct
from the effects of dust the abundance ratios
[Si/Fe], [S/Fe], [Ti/Fe], [Mn/Fe], [Ni/Fe] 
available for a sub-sample of 9 absorbers.  
The  [$\alpha$/Fe] ratios corrected from dust do not show   
the enhancement characteristic
of metal-poor Galactic stars, but instead
have essentially solar values, within $\pm 0.2$ dex.  
This suggests that
the chemical history of DLA absorbers is different from
that experienced by the Milky Way.
Evidences that point to dwarf galaxies,
rather than to spiral galaxies, as important contributors
to the DLA phenomenon are summarized.

\end{abstract}
 
\keywords{QSO absorption systems --- 
Damped Ly $\alpha$ absorbers --- 
galaxies: abundances, evolution }

\section{Introduction}

Damped Ly\,$\alpha$ (DLA) absorbers are the class
of QSO intervening systems with highest hydrogen column
density, $N_\HI \geq 2 \times 10^{20}$ atoms cm$^{-2}$.
Several properties of these absorbers indicate
that they are high-redshift
progenitors of present-day galaxies 
(Wolfe et al. 1986, Wolfe et al. 1995). 
Spectroscopic observations show, in addition to high 
HI column densities,  metal lines of low ionization and      
velocity dispersions typical of gas rich galaxies. 
Imaging studies of fields around QSOs with foreground
DLA absorption at low redshift ($z \leq 1$)
reveal the presence of gas-rich galaxies with a variety
of morphological types (Le Brun et al. 1997). 
An indepedent and remarkable piece of evidence in favour
of a galactic origin comes from the behaviour of $\Omega_g(z)$,
the comoving
mass density of neutral gas in DLA absorbers in units
of the critical density of the Universe. 
The fact that $\Omega_g(z=3.5)$
equals the mass density of luminous matter (stars)
at the present epoch, and that
$\Omega_g(z)$ decreases with time,
is interpreted as evidence of
gas consumption due to star formation since
$z=3.5$ to the present time (Wolfe et al. 1995). 
At high redshift  $\Omega_g(z)$ 
is at least 10\% of the barionic content of the universe
and therefore understanding the nature of DLA galaxies
will give important contraints
to cosmological models involving structure formation  
in the early universe
(Klypin et al. 1995).  
A commonly accepted origin of DLA absorbers is in spiral galaxies,
(Wolfe et al. 1995, Prochaska \& Wolfe 1997b),
but an origin in dwarf galaxies has also been
suggested (York et al. 1986; Tyson 1988; 
Pettini, Boksenberg \& Hunstead 1990;
Molaro et al. 1996; Matteucci, Molaro \& Vladilo 1997).  
 
Studies of elemental abundances in DLA systems
give important clues
for understanding which kind(s) of galaxies are 
responsible for the absorption. 
In addition, the study of DLA
abundances  over a large redshift interval
will eventually allow us to trace 
the chemical evolution of the absorbers 
over a large interval of look-back time, up to the early
stages of galactic  nucleosynthesis,
providing  fundamental tests to models of galactic chemical
evolution. 

Abundance measurements in DLA systems  do not depend 
critically on ionization corrections because
these absorbers are optically thick to ionizing
radiation (h$\nu > 13.6$ eV). As a consequence,  
species that are dominant in HI regions have negligible
ionization corrections, as confirmed by detailed photoionization
calculations (Lu et al. 1995; Viegas 1995). 
Nevertheless, the abundances observed in DLA absorbers
may not represent the
real chemical composition of the system if part of the elements is 
removed from the gas to the solid phase (dust grains), 
as it happens in the interstellar medium of our Galaxy where 
elements with very different dust depletions 
 are known to exist  
(Jenkins 1987; Savage \& Sembach 1996). 
In DLA absorbers the relative abundances of elements expected to
have different  dust depletions, such as Zn and Cr,   
suggest that dust is indeed present, even though
with a lower dust-to-gas ratio than in the Galaxy
(Pettini et al. 1994, 1997).   More
direct evidence for dust in DLA systems has been found from
 reddening studies of QSOs with and without foreground 
damped absorption (Pei, Fall \& Bechtold, 1991). 

A relatively large number of abundance determinations
in DLA absorbers has been collected  
since it has become possible to observe QSOs at
sufficiently high spectral resolution
(see list of references quoted in Table 2).
Attempts to interpret the relative abundances
(i.e. element-to-element abundance ratios)
observed in DLA absorbers have lead, however, to contradictory
conclusions.
From one side claims have been made that the pattern
of relative abundances 
resembles that of Galactic halo stars,
having in common an enhancement of the alpha elements 
and a deficiency of manganese relative to iron 
(Lu et al. 1996, Prochaska \& Wolfe 1997).
From the other side, evidence has been presented for nearly solar
[$\alpha$/Fe] ratios in spite of the low
metallicity level of the absorbers 
(Molaro et al. 1996, Molaro et al. 1997). 
These contradictory conclusions are probably due to the fact that 
the presence of dust in DLA absorbers is not properly taken into account.  
Several attempts have been done to reproduce the observed abundances
by combining different intrinsic abundance patterns 
with dust depletion patterns of Galactic interstellar  type
(Lauroesch et al. 1996; Kulkarni, Fall \& Truran 1997).
According to these studies, neither a halo-like nor a
solar-like metallicity pattern match the observations perfectly. 
In the present work the effect of the dust
on the determinations of the elemental abundances 
is studied in detail by deriving equations that take into account 
the level of metallicity, the dust-to-gas ratio
and the intrinsic abundance pattern of DLA absorbers. 
Differently from the analysis performed by
Kulkarni, Fall \& Truran (1997), in the present study
each individual DLA system is treated separately and 
the extinction induced by the dust
is taken explicitely into account. 
The general equations that
link observed abundances to real abundances
of the absorbers are presented in Section 2. 
Simplified forms of these equations are 
then used in Section 3 to predict 
possible dust depletion patterns in DLA systems. 
In Section 4 is presented the method for estimating dust-to-metals ratios,
dust-to-gas ratios and dust-corrected abundances
in individual DLA absorbers;
the method is applied to available
abundance measurements of  DLA systems. 
The relations between dust and metals in DLA absorbers
are examined in Section 5.  
The implications for our understanding of the nature of DLA galaxies are discussed
in Section 6 and the results are summarized in Section 7.

\section{Definitions and basic equations}
 
The observed abundance ratio by number (X/H)$_\obs$  of an element  X  
in a DLA absorber
is  determined from measurements of the  column densities
 in the  gas phase of the system,  $N_\Xg$ and $N_\Hy \simeq N_\HI$.  
The observed abundance  is
usually expressed in logarithmic form relative to solar composition
\begin{equation}
\left[ { \X  \over \Hy } \right]_\obs 
= \log ( {  \X \over \Hy } )_\obs   - 
  \log ( {  \X \over \Hy } )_{\sun} 
~.
\end{equation}  
However, this is not the real abundance since  
the column density
of atoms  in dust, $N_\Xd$,
is not considered in (1). 
The total column density ($N_\Xg+N_\Xd$)
 would instead give the
real ratio (\X/\Hy)$_\intr$ and hence the real abundance
\begin{equation}
\left[ { \X \over \Hy } \right]_\intr 
=  \log ( {  \X \over \Hy } )_\intr      - 
                  \log ( {  \X \over \Hy } )_{\sun}
~.
\end{equation}
By introducing the {\em fraction in dust}  of the element X
\begin{equation}
f_\X =  N_\Xd / (N_\Xg + N_\Xd)  
\end{equation} 
it is easy to derive a relation between the observed and the real abundance
\begin{equation}
\left[ { \X \over \Hy } \right]_\obs  
= \left[ { \X \over \Hy } \right]_\intr  + 
\log  \left(  1 - f_{\X}  \right)  
~~.
\end{equation}

To introduce the effect of the dust in the above expression  we start
by defining the  dust-to-gas ratio
\begin{equation}
k =   \tau  / N_\HI  
\end{equation}
where $\tau$ is the extinction optical depth
in the spectral region of interest in the rest frame of the absorber.
The above definition is the same of Pei, Fall \& Bechtold (1991) 
provided the hydrogen column density$N_\HI$ is expressed in units of 10$^{21}$ cm$^{-2}$. 
If we call $N_{\rm gr}$  the column density of dust grains 
per unit area along the line of sight, the optical depth is given by
\begin{equation}
\tau  =   N_{\rm gr} \:  c_{\rm e} 
~,
\end{equation}
where  $c_{\rm e}$ is the {\em average extinction cross-section of the grains}.
For spherical  grains of radius $a$ the cross-section for extinction
is $C_{\rm e} = \pi \:  a^2 \: Q_{\rm e}$  where
the efficiency factor for extinction $Q_{\rm e}(\lambda)$
can be estimated in the framework of Mie scattering theory 
(see e.g. Wickramasinghe 1967). 
If we call $n_{\rm gr}(a) \, {\rm d}a$
the  number of grains  with
radii in the range ($a$,$a+{\rm d}a$) per unit volume, then it is possible
to define $c_{\rm e} = < \pi \:  a^2 \: Q_{\rm e} >$  through the relation
$
\tau  =  \int \pi \:  a^2 \: Q_{\rm e} \: n_{\rm gr}(a) \: {\rm d}a
= N_{\rm gr} \: c_{\rm e} 
~.
$
We will use, however, the most general form (6) to avoid
any assumption on the geometry  of the grains.  

We then define
  the {\em average number of atoms of the element X per dust grain}  
\begin{equation}
\nu_\X  = N_\Xd / N_{\rm gr}
\end{equation}
which is related to the chemical composition of the dust
since the   number ratio of two elements X and Y
in the dust  grains will be
(X/Y)$_{\rm gr}$ = $\nu_\X / \nu_\Y$.

From (5) , (6) and (7) it is easy to see that
$  f_\X =   k   (   \X         /  \Hy      )_\intr^{-1}  
           (   \nu_\X / c_{\rm e} ) $, 
where  $   (\X/\Hy)_\intr = ( N_\Xg + N_\Xd ) / N_\HI$. 
By defining the metallicity level of the system
as $Z = (\Y/\Hy)_\intr$, where Y is a suitable reference element,
we obtain: 
\begin{equation}
f_\X 
=          { k   \over    Z                 }  
           \left( { \X          \over  \Y      } \right)_\intr^{-1}  
            {   \nu_\X   \over c_{\rm e} }
            ~.
\end{equation}

From (4) and (8) we obtain a relation between observed and real abundance:
\begin{equation}
\left[ { \X \over \Hy } \right]_\obs  
= \left[ { \X \over \Hy } \right]_\intr  +
\log  
\left(                                                       
1-   {k    \over     Z                  }   
    \left( {\X            \over    \Y                 } \right)^{-1}_\intr   
             {   \nu_\X   \over c_{\rm e} }  
                                                                  \right) 
\end{equation}
where  the contributions
of dust-to-gas ratio, metallicity level, elemental abundance pattern,
dust chemical composition and dust extinction are clearly separated.  
A crucial role in the relation is played by the {\em dust-to-metals ratio} $(k/Z)$.

\subsection{Scaling to Galactic quantities}

Relation (9) is not suited for comparing observed and real abundances
because it includes several unknown quantities.   
In order to obtain a more useful relation 
it is convenient  to normalize the quantities in the   DLA absorbers
in units of the corresponding Galactic quantities. 
We indicate such normalized values with a tilded symbol, 
e.g. $\ktilde = k / k_\Gal$. 
By applying (8) to DLA systems  and to  our Galaxy, we derive
\begin{equation}
f_{\X} = {  \ktilde     \over       \Ztilde       }      \cdot
             10^{  -\left[ {\X \over \Y} \right]_\intr  }        
                                                          \cdot
        {   \nutilde_\X   \over \ctilde_{\rm e} }          \cdot 
f_{\X,\ism}
~, 
\end{equation}
where 
(i)  $\Ztilde = Z/Z_{\sun} = 10^{  [\Y/\Hy]_\intr  }$,
(ii) $10^{  \left[ {\X \over \Y} \right]_\intr  } =
      (\X/\Y)_\intr / (\X/\Y)_{\sun}$
since we adopt solar abundances for the Galaxy,
and 
(iii) the  fraction in dust   $f_{\X,\ism}$ refers to the 
Galactic interstellar medium. 
 
This expression for the     fractions in dust
can be written in a simpler form if  
 we require the condition  
$\nu_\X = \nu_{\X,\ism}$ for any element,
which is equivalent to assume that
{\em the dust has same chemical composition 
and same average number of atoms per   grain
as Galactic dust.}
 This equivalence follows   from the fact that
the relative abundance of two elements in the dust is
$(\X/\Y)_{\rm gr}  = (\nu_\X/\nu_\Y)$. A constant relative
abundance in the dust is
consistent with current understanding of grains chemical composition
(see e.g. discussion in Section 12.2 and 12.3 by Savage \& Sembach 1996).
In practice, once the chemical composition and the average number of atoms
are specified, also the grain sizes are specified. 
Therefore, also the 
extinction coefficient, which depends on the
dielectric properties and geometry of the grains  is  determined,
the only left degree of freedom being the exact shape of the grains. 
Therefore, once the   condition $\nu_\X = \nu_{\X,\ism}$ is satisfied, also
condition $c_{\rm e} = c_{{\rm e},\ism}$ is virtually satisfied, 
and hence 
$\nutilde_\X   = \ctilde_{\rm e} = 1$.
Relation (10) becomes
\begin{equation}
f_{\X} = {  \ktilde     \over       \Ztilde       }      \cdot
             10^{  -\left[ { \X \over \Y } \right]_\intr  }                  
                                                         \cdot
f_{\X,\ism}
\end{equation} 
and from (4)  we obtain the equation
\begin{equation}
\left[ { \X \over \Hy } \right]_\obs        = 
\left[ { \X \over \Hy } \right]_\intr  + 
\log  
\left(  1 -  {  \ktilde \over \Ztilde  }  \cdot
 10^{  - \left[ { \X \over \Y } \right]_\intr  }  
 \cdot  f_{\X,\ism} \right)  
\end{equation} 
that can be used
to estimate the gas-phase abundances  given the real
metallicity pattern  and a suitable combination 
of dust-to-gas ratio $\ktilde$, and metallicity $\Ztilde$.  
The   fractions in dust $f_{\X,\ism}$ can be considered as input parameters
which are
known with reasonable accuracy from Galactic interstellar studies,
as we show in the next section.

\section{Dust depletion pattern in DLA systems}
 
By analogy to the definition of  logarithmic dust depletion in the 
Galactic interstellar medium
\begin{equation}
\delta_{\X,\ism} =  \log ( {  \X \over \Hy } )_\obs          - 
                    \log ( {  \X \over \Hy } )_{\sun}
\end{equation}
we define the dust depletion   in DLA systems as
\begin{equation}
\delta_{\X,\dla} =  \log ( {  \X \over \Hy } )_\obs           - 
                    \log ( {  \X \over \Hy } )_{\intr}
\end{equation}
where the number ratios $( \X / \Hy)_\obs$ are measured in the gas
phase in each case. 
Since dust  depletion and   fraction in dust (3) are related
by the equation
\begin{equation}
\delta_\X =  \log (1 - f_\X  )
\end{equation}
it follows from (11) that for Galactic-type dust
\begin{equation}
\delta_{\X,\dla}   = 
 \log  
\left(  1 -  {  \ktilde \over \Ztilde  }  \cdot 
                        10^{  -\left[ { \X \over \Y} \right]_\intr  }  
\cdot   f_{\X,\ism} \right)  
.
\end{equation}
We have used this relation to estimate  dust depletion patterns
in DLA systems  by using the  dust-to-metals ratio
$(\ktilde / \Ztilde)$ and 
the   abundance pattern $[\X/\Y]_\intr$
as free parameters. The results are shown
in Table 1, where elements most commonly measured in DLA systems 
are considered. Iron is used as the zero point  for the
metallicity level. 
We have computed  the
  fractions in dust $f_{\X,\ism}$   from  (15)  by adopting 
dust depletions typical of Galactic warm disk  
(Savage \& Sembach 1996). More details are given in
the notes to Table 1.  

The results in  the columns labeled S100, S75, and S50
 are relative to a solar 
  abundance pattern ($ [ \X / \Fe ]_\intr = 0  $) for dust-to-metals
ratios  $(\ktilde / \Ztilde )$ = 1.00, 0.75 and 0.50, respectively.  
By definition, the case S100 is nothing else than the Galactic
depletion pattern for low density lines of sight (Galactic warm disk).
By comparing this pattern with the cases S75 and S50
one can see that the absolute values of the depletions decrease
when the dust-to-metals ratio is reduced. 
However, 
{\em the scaling law
of the differential depletion
is not linear}, contrary to what assumed in previous works
(see e.g. Lauroesch et al. 1996).  
Rather,  
{\em the differential depletion between pairs of elements
tends to vanish as the dust-to-metals ratio decreases}.
This implies that
elements   moderately depleted in the Galaxy can still have
a non negligible depletion when less dust is present. 
As an example, 
Si depletion decreases by only 0.3 dex, compared to a decrease
of 1 dex of Fe,  
when   $(\ktilde / \Ztilde )$ is reduced from 1.0 to 0.5.

More surprising results  are found when an  
abundance pattern different from the solar one is considered. 
In the columns labeled H100, H75, and H50 
we give the depletions   expected for an intrinsic abundance pattern 
typical of Galactic halo stars, again  for dust-to-metals
ratios  $(\ktilde / \Ztilde )$ = 1.00, 0.75 and 0.50, respectively.  
The adopted values of
halo star abundances   
are taken  from Francois (1988), Sneden, Gratton \& Crocker (1991),
and Ryan et al. (1996),  
at a metallicity level $Z/Z_{\sun} = 0.1$
 characteristic of DLA absorbers; 
we refer to the notes  
in Table 1 for more details. 
The very small deviations from the solar  pattern found for [Cr/Fe]
and [Ni/Fe] 
are treated here as real   to show how even   small deviations
can affect the dust depletion pattern. 
From the results shown in Table 1 
one can see that 
{\em  inversions in the differential depletion pattern
may appear if the DLA abundance pattern is not solar.}
For instance, underabundant elements such as
Mn and Cr are more depleted than Fe in halo-like models,
contrary to what found in  Galactic interstellar gas. 
Overabundant elements tend, on the other hand, to be less depleted.
As an example, Ti depletion is quite modest in halo-like models
while it is comparable to iron depletion in Galactic gas. 
It is also interesting to note that
{\em no solutions are allowed for underabundant elements  
with high   fraction in dust ($f_{\X,\ism} \simeq 0.9$)
unless the dust-to-metals ratio is sufficiently low.}
This apparently peculiar result is due to the fact that
 one cannot  use {\em whatever} combination of
input parameters because by definition (3)  must be  
$f_{\X} \leq 1$  and therefore  
$      ( \ktilde   /  \Ztilde  )      \leq  
(   10^{  [\X/\Y]_\intr  } / f_{\X,\ism} )  $,
otherwise there will be not enough atoms available to form the 
required amount of dust.
As an example, there are no solutions for Mn in the models H100 and H75,
while there is solution for H50, when the dust-to-metals ratio becomes
half the Galactic value.

\section{Analysis of abundance measurements}

Instead of using Eq. (12) to estimate the effect of a given
DLA abundance pattern on the observed abundances, we
will follow the opposite approach   of deriving
the real abundance pattern
starting from the observed   ratios.
By applying Eq. (12) to the elements X and Y and
combining the resulting equations it is possible to find 
the real abundance ratio
\begin{equation}
10^{     \left[  {\X \over \Y}  \right]_\intr   }      =
10^{     \left[  {\X \over \Y}  \right]_\obs         } +
     {  \ktilde \over \Ztilde  }   \cdot   
\left(  
   f_{\X,\ism}   -   
    f_{\Y,\ism} 
\cdot 10^{  \left[ { \X \over \Y } \right]_\obs }  
\right)
\end{equation} 
from  the observed abundance ratio.
This equation 
still depends on the dust-to-metals ratio which is unknown. However,
 if two elements X and Y trace each other
over the chemical history of the system,
i.e. if $ [ \X / \Y ]_\intr = 0  $ at any time,
then  it is easy to derive from (17) the expression
\begin{equation}
 {  \ktilde \over \Ztilde  }   = 
                                                              { 
 10^{     \left[  {\X \over \Y}   \right]_\obs   } - 1    
                                                              \over
 f_{\Y,\ism} \cdot 10^{ \left[  {\X \over \Y}  \right]_\obs }  
 -   f_{\X,\ism}  
                                                              }     
\end{equation} 
that allows to compute the dust-to-metals ratio from observed
abundances   only, by
using the Galactic interstellar   fractions in dust as input parameters.
Once  the dust-to-metals content is determined from a given pair
of elements, 
Eq. (17) can be applied to other elements for deriving
their intrinsic abundance.
This method can be applied to individual
DLA systems for which
suitable abundances measurements are available. 
In the following sub-sections we describe the results of 
the application of this procedure to currently
available measurements.

\subsection{Dust-to-metal ratios from iron-peak elements}

In order to determine the dust-to-metals ratio from
Eq. (18) we need at least two elements for which
$ [ \X / \Y ]_\intr = 0  $ in  DLA systems. 
Studies of metal poor stars in the Galaxy indicate that
iron-peak elements elements generally trace each other 
down to metallicities comparable to those found in DLA absorbers
(see third column of Table 1 and references quoted therein). 
A relatively large amount
of observations of iron-peak elements are available for DLA
systems, as can be seen in the updated summary of measurements
  presented in Table 2. 
The use of iron-peak elements is therefore the most natural 
choice to determine the dust-to-metals ratio. 
However, a few caveats have to be considered.  
From one side,  observed DLA [Mn/Fe] ratios  
show an underabundance which is difficult to understand
in terms of dust depletion  (Lu et al. 1996). 
In addition, some deviations from solar  ratios are
found in metal-poor stars
for [Cr/Fe], [Mn/Fe], [Co/Fe] (Ryan et al. 1996), 
and for [Cu/Fe] (Sneden, Gratton \& Crocker 1991). 
For Cr, however, these deviations are significant only at
[Fe/H] $< -2.5$ and are negligible
at the metallicity level typical of DLA systems. 
Taking also into
account the number of available  measurements, 
this leaves Zn, Fe and Cr as the best candidates for using Eq. (18).
Thanks to the long-standing observational effort of 
Pettini et al. (1994, 1997),
a relatively large number of Zn and Cr measurements are available.
These overlap in many cases with Fe (and sometimes Ni) determinations 
based on Keck observations (Lu et al. 1996 and refs. therein).
The form of the denominator of the right side of (18) indicates
that the use of any pair of elements with similar fractions in
dust and
gas-phase abundances
must be avoided, because a small observational error
would lead to a large error in the   dust-to-metals ratio.
This precludes the use of Cr and Fe, as can be seen in  
Tables 1 and 2. Since Zn has a
much smaller fraction in dust than  Cr and Fe 
we have determined   the dust-to-metals ratio from (18)
by using Zn as the  
element X and Cr or Fe as  the element Y.  

Table 2 includes  the
    17 DLA systems for which a Zn measurement exists,
rather than an upper limit, together
with a Cr or Fe determination. 
The resulting dust-to-metals ratios are shown in Table 3.
In 8 cases the determinations
are obtained independently from  [Zn/Cr]   
and   [Zn/Fe] ratios. We do not find a systematic
difference between the 8 results in common, the average  
percent difference being $-0.06 \pm 0.19$. 
This agreement  
gives us confidence on the method
adopted and on the assumption that these elements indeed 
trace each other  
in DLA systems, at least down to the level of metallicity
of the present sample. 
For these 8 cases we adopt  
the average of the two independent determinations as the
final dust-to-metals ratio. 
For the other cases we adopt the individual result,
being confident that Fe vs. Cr systematic effects should be negligible.
An attempt to derive the dust-to-metals ratio using [Zn/Ni]
leads to values systematically higher by about 20\% than the values
derived from [Zn/Cr] and [Zn/Fe]. This suggests that a weak deviation
from solar iron-peak relative composition may be present for Ni,
even though the suspect has been advanced that Ni abundances
could be affected by uncertainties in the NiII oscillator strengths
(Lu et al. 1996). 
For these reasons,  dust-to-metals ratios determined from  [Zn/Ni] ratios
are not considered here. 

The typical random error of the dust-to-metals ratio 
derived in this way can be estimated  
by neglecting the contribution of $f_{\Zn,\ism}$ in the denominator
of the right side of (18). The error
of the dust-to-metals ratio equals that of the [Zn/Y] measurement,
which is typically of 0.1 dex, although
in some cases it can be as low as 0.05 dex.   
In addition to this random error, systematic errors are introduced
through the Galactic fractions in dust used in (18). 
To quantify this effect we computed the dust-to-metals in two cases:
(i) zinc completely undepleted, i.e. $f_{\Zn,\ism} = 0$, and
(ii) zinc mildly depleted, with  $f_{\Zn,\ism} = 0.35$.  
The resulting dust-to-metals ratios, also 
shown  in Table 3, are systematically lower by $\simeq$ 20\%
in the first case relative to the second case. 
Uncertainties in Fe and Cr  fractions in dust
would yield a much smaller systematic error since 
$f_{\Fe,\ism} \simeq f_{\Cr,\ism} \simeq
0.9$ for current range of Fe and Cr dust depletions 
observed in Galactic interstellar clouds.

\subsection{Relative abundances corrected from dust effects}

Having derived the $(\ktilde/\Ztilde)$ ratio for individual
DLA systems we can now use  relation (17) to estimate  
relative  abundances    corrected from dust effects. 
In Table 4 we list additional abundance measurements  
available for the DLA systems of the sample shown in Table 3.
Dust-corrected relative abundances are given in Table 4 below each
gas-phase determination and are shown in a synoptic view in Fig. 1.  
Internal errors of the corrected relative abundances are quite
low, as estimated from the dispersion of the 5 [Si/Fe] measurements,
$\pm 0.09$ dex, and of the 7 [Ni/Fe] measurements,
$\pm 0.04$ dex. 
By applying   error propagation to (17) one can see that
the total random error of the dust-corrected ratios
is typically $\simeq$ 0.15 dex, but can be as low as
$\simeq$ 0.05 dex for the most precise measurements. 
In addition, systematic errors related to the uncertainty of the
$f_{\X,\ism}$ and $f_{\Y,\ism}$     parameters should be
taken into account. 
The use of dust-to-metals ratios derived by assuming zinc
completely undepleted would systematically   reduce the dust corrections;
however, the net effect on the corrected ratios is in most cases
less than  0.05 dex. 

A first result clear from Fig. 1 is that
the spread of the dust-corrected relative abundances 
for any given [X/Fe] ratio is generally within
the range of the expected errors, suggesting  that 
{\em  the sample of DLA systems investigated is chemically homogeneous 
over the range of metallicities from nearly solar
down to --1.5 dex. }  

The most important result visible if Fig. 1, however,
comes from the analysis of the element-to-element abundance pattern. 
If the intrinsic abundances
of DLA systems were similar to metal-poor   Galactic halo
stars, the ratios shown in Fig. 1 should deviate
from the solar pattern, with the only exception of [Ni/Fe].  
Quite surprisingly 
{\em the dust-corrected ratios are      
remarkably close to solar.}   
Errors due to uncertainties of the fractions in dust
$f_{\X,\ism}$ and $f_{\Y,\ism}$ should produce different effects
for any given pair of elements X and Y, and would tend
to destroy the solar-like pattern. 

Of course, as we discuss in detail in the following paragraphs, the
relative abundances are not {\em exactly} solar. 
It is also important to note here that  
departures from solar ratios in DLA systems are found 
when nitrogen abundances are considered
(Lu 1997, priv. comm.; Vladilo et al. 1997b). 
However, owing to the complex galactic chemical
evolution of this element, the study of nitrogen abundances
is deferred to a subsequent paper
(Centuri\'on et al. 1997). 
We discuss now in more detail specific (groups of) elements. 

\subsubsection{Nickel}

Nickel is expected to trace iron at least down to  
a metallicity $Z/Z_{\sun} \simeq 0.01$, even though there
is some indication for a very mild underabundance
($\simeq -0.02 $ dex) at $Z/Z_{\sun} \simeq 0.1$
from the data by Ryan et al. (1996). 
The 7 dust-corrected measurements available yield
$< [\Ni/\Fe] > = -0.06 \pm 0.05$, consistent both with
Ni   tracing Fe or with a mild underabundance.
Even if this result is not particularly surprising, it shows already
the importance of the correction from dust effects.
In fact the gas phase [Ni/Fe] measurements yield values
around $-0.2$ dex, which have lead Lu et al. (1996)
to suspect the presence of a constant offset between
stellar and DLA measurements. Our results can be instead explained
  in terms of the expected trends of iron-peak elements.   

\subsubsection{Alpha elements}

The ratios of alpha elements to iron-peak elements is the best
indicator of halo-like metallicity pattern.
In the early stages of chemical evolution the products of type II
SNae, rich in alpha elements, are expected to dominate the observed
abundances and to produce the overabundance of
the [$\alpha$/Fe] ratio which is observed in Galactic metal-poor
stars. At later stages, the contribution of type Ia SNae 
  reduces the [$\alpha$/Fe] ratio to the typical
values observed in Galactic   stars of solar metallicity
(see e.g. Matteucci, Molaro \& Vladilo 1997). 
At a metallicity   $Z/Z_{\sun} \simeq 0.1$
Galactic halo stars show [Si/Fe] =+0.27 dex (Ryan et al. 1996)
and [S/Fe] = +0.4 dex (Francois 1988; including a correction of $-0.2$
dex proposed by Lambert 1989). 
Ti behaves similarly to an
$\alpha$ element with  [Ti/Fe] = +0.22 at 
$Z/Z_{\sun} \simeq 0.1$ (Ryan et al. 1996). 
If the chemical history of DLA systems were similar to that of the
Galaxy, we would expect to find 
similar overabundances of the [$\alpha$/Fe] ratios. 
This is not the case for any of the elements considered
here. 
The 5 dust-corrected measurements of Si   yield
$< [\Si/\Fe] > = +0.02 \pm 0.09$ (internal error of the average).
Even assuming a total random error as high as 0.15 dex,
the result is still inconsistent at two sigma level
from the [Si/Fe]  $\simeq$ +0.3 dex
expected for a halo-like abundance pattern. 
Also the 3 dust-corrected measurements of [S/Fe] do not show   
evidence of  overabundance. 
As noted by Molaro, Centuri\'on \& Vladilo (1997),
in all these 3 cases the observed  [S/Zn] ratio  is solar.
Since dust depletion is modest for both S and Zn
this result confirms the validity of the correction applied to the
observed [S/Fe] ratios. 
Of the two dust-corrected measurements of  [Ti/Fe], one is
clearly not overabundant and the other is consistent, within
an error bar of $\pm 0.15$ dex both with a solar-like
abundance and with the mild overabundance observed in
Galactic halo stars. 
Finally, we note that also
the  [Mg/Fe] upper limit derived by Molaro, Centuri\'on
\& Vladilo (1997)
for the $z = 2.309$ absorbers toward PHL 957
is inconsistent with a halo-like pattern.

\subsubsection{Manganese}

Gas-phase measurements of  the [Mn/Fe] ratio in DLA systems
systematically yield negative values (Lu et al. 1996). 
The [Mn/Fe] ratio is believed to be a good discriminant between  
Galactic halo and Galactic interstellar abundance patterns.
In the first case Mn is underabundant 
([Mn/Fe] = $-0.35$ dex; Ryan et al. 1996)  
while in latter case Mn is overabundant 
( [Mn/Fe] $\simeq$ $+0.3$; Savage \& Sembach 1996)
as a consequence of differential dust depletion. 
However, as we have shown in Section 3, even a mild  
underabundance of the [Mn/Fe] ratio can lead to an inversion
of the dust depletion pattern, manganese becoming more depleted
than iron. This explains why the dust correction can increase
the observed [Mn/Fe] ratio, 
as it happens in the two cases shown in Table 4.
The dust corrected ratios   indicate in fact a modest
underabundance of manganese. This underabundance 
is smaller than expected for a halo-like chemical composition
and the [Mn/Fe] ratio is marginally consistent with a solar composition. 
We note that this conclusion does not depend on the
adopted cosmic abundance for manganese. For consistency
with the results of Ryan et al. (1996) we have renormalized
the DLA measurements by adopting 
solar photospheric abundance for Mn, instead of the solar meteoric
value used by Lu et al. (1996). Should we adopt
the meteoritic value, we would find a [Mn/Fe] lower by
0.14 dex in the DLA systems,
but also the    [Mn/Fe] ratio for Galactic halo stars should
be   decreased, down to [Mn/Fe] = $-0.49$ dex, and
the dust corrected ratios would still be in between the solar-like
and the halo-like case.

\subsection{Dust-to-gas ratios}

The basic assumption of the present work is that
dust affects the observed abundances in DLA absorbers. 
We can test this assumption by comparing the dust-to-gas ratios
derived from the observed abundances with dust-to-gas ratios obtained
from reddening determinations. 
For any two elements X and Y with similar nucleosynthetic history 
it is possible to estimate the dust-to-gas ratio
by means of the expression
\begin{equation}
\ktilde = 
                                            { 
 10^{     \left[  {\Y \over \Hy}  \right]_\obs   } - 
 10^{     \left[  {\X \over \Hy}  \right]_\obs   }  
                                                     \over
    f_{\X,\ism} -    f_{\Y,\ism}
                                           }     
\end{equation} 
which can be derived by applying (12) to X and Y
and by combining the resulting relations with the condition 
  $ [ \X / \Y ]_\intr = 0 $.
This method is conceptually similar to that  
originally used by Pettini et al. (1994)
for estimating the dust-to-gas ratio  
from Cr and Zn abundances. 
 
On the basis of what discussed in Section 4.1
we have used the  [Zn/H] and [Cr/H] ratios,
and the [Zn/H] and [Fe/H] ratios, 
to derive  $\ktilde$ from (19)
for our sample of 17 DLA systems.
The results are shown in Table 3. 
In the 8 cases with available independent determinations
we do not find a systematic
difference between the results based on the [Zn/H] and [Cr/H] ratios
and the results based on the [Zn/H] and [Fe/H] ratios.
The average  percent difference between the two cases
is $-0.07 \pm 0.23$. By applying random error propagation to (19),
keeping fixed the Galactic fractions in dust
$ f_{\X,\ism} $ and $ f_{\Y,\ism}$, 
one finds about twice larger errors than in the case of the relation (18).
This is not surprising since two independent measurements
are required to determine   $\ktilde$ from (19)
and only one to determine   $(\ktilde/\Ztilde)$ from (18).
Systematically lower dust-to-gas ratios by about $-$0.2 dex
would be derived by assuming zinc completely undepleted.

The resulting dust-to-gas ratios  
span a wide interval, from about 1\% to 75\%  the Galactic value,
without a concentration around a particular value. 
The logarithmic average yields
$<\log \ktilde> = -1.12 \pm 0.52$, which implies that the most probable
interval lies between  2\% and 25\% the Galactic value. 
This result is in good agreement with the most probable range    
found by Pei, Fall \& Bechtold (1991) in
their study of QSOs with and without foreground DLA absorbers,
which is between  5\% and  20\%. 
The combined sample used by these authors includes 22 QSOs 
with foreground DLA absorber, 7 of these systems being in common 
with the present sample. 
The quantitative agreement between the two   independent
extinction estimates   supports   the validity of the 
assumption that dust is responsible for the apparent deviations
of the [Zn/Cr] and [Zn/Fe] ratios from solar composition. 
By assuming zinc completely undepleted, our results
indicate a most probable range  from  1\% up to  16\%  of the Galactic value.
Even if this interval still overlaps with the best range
of Pei, Fall \& Bechtold (1991), the mild zinc
depletion adopted here yields a better agreement.

\subsection{Metallicities corrected by dust effects} 
 
Once we have determined the dust-to-gas ratio,
the metallicity corrected by dust can be determined
from the relation
\begin{equation}
\Ztilde = 
10^{     \left[  { \Y \over \Hy }  \right]_\obs            }    +
\ktilde   \cdot    f_{\Y,\ism}
\end{equation}  
which results by applying Eq. (12) to the element used as
the reference for the metallicity level. 
 The dust-corrected metallicities of our sample are shown in Table 3.  
Since Fe, Cr and Zn  trace each other, 
equivalent results are obtained by using any of the
three elements in Eq. (20). In practice we have used the 
[Zn/H] ratios for which the correction term is smaller.
On the average
the corrected metallicities   differ by only $\simeq$ +0.12 dex
from the observed [Zn/H] ratios, which would  represent
the real metallicity if zinc were completely undepleted.  
Owing to the small amount of the correction, the use of the 
dust corrected metallicities of Table 3
does not affect significantly the values of 
metallicities derived for DLA absorbers.
It is worth noting, however, that even a small amount of zinc
depletion in the Galaxy still implies a small but measurable
amount of depletion in DLA absorbers. 
This can be understood from our previous discussion in Section 3.

\section{Correlation between dust and metal content in DLA absorbers}

An important by-product of the present investigation is the
determination of dust-to-gas ratios and metallicities
in individual DLA systems. 
We start by analysing the dust-to-metals ratio $\ktilde/\Ztilde$,
which is less affected by error propagation than
the dust-to-gas ratio $\ktilde$. 
The logarithmic average yields $< \log (\ktilde/\Ztilde) >
= -0.21 \pm 0.16$, which implies a typical value around 62\%
and the most probable range between 42\% and 89\% the Galactic
dust-to-metals content. 
An attempt to find correlations of  $(\ktilde/\Ztilde)$
with metallicity or with
redshift gives negative results. We conclude that
in the present sample of 17 DLA systems 
{\em the dust-to-metals ratio is approximately constant
and   lower than in the Galaxy.}  
The average dust-to-metals ratio that we find is  
comparable with the value $\ktilde/\Ztilde \simeq 50\%$   obtained 
by Pettini et al. (1997) by assuming zinc completely undepleted. 

The constancy of the $\ktilde/\Ztilde$  ratio 
implies that dust-to-gas ratio and metallicities are
correlated in DLA systems. 
This can be seen in Fig. 2, where we  plot  
$\ktilde$ versus metallicity $\Ztilde$ for our sample
of 17 DLA systems. Filled diamonds represent  
the 8 systems with most accurate measurements,
based on two independent $\ktilde$ determinations in each case.
The other systems, with one single $\ktilde$ determination
are indicated with empty diamonds. 
The linear regression through the 17 DLA data points  
confirms the existence of a correlation,
with slope  1.16 and  correlation coefficient $r=0.95$
(dotted line in the figure).
By using only the 8 systems with most accurate measurements
we find a correlation with 
slope 0.98 and $r=0.95$ (dashed line). 
The slope is therefore 1 within the errors, as expected for a 
constant $\ktilde/\Ztilde$  ratio. 

The fact the dust content follows the metal content 
gives fresh support to the  idea
that DLA systems are originated in galaxies:
dust would be produced in the diffuse medium following
the injection of  metals from the stellar component. 
The constancy of the $\ktilde/\Ztilde$  ratio
  has important implications
in the framework of the model of cosmic chemical evolution developed
by Pei \& Fall (1995). In fact, in that model 
{\em it is assumed} that
the dust-to-metals ratio is constant in DLA absorbers.  
Our study {\em shows}
that this basic assumption of the model  is essentially correct. 
The range of redshifts investigated here is probably insufficient
to detect variations of the dust-to-metals content that may
be expected in the course of the cosmic evolution.

It is worth noting  that the constancy of the dust-to-metals ratio
could have been predicted from the remarkably
small scatter of the abundances measured in different
DLA systems. In fact, 
each one of the ratios [Si/Fe], [Cr/Fe]  and [Ni/Fe] from
the compilation by Lu et al. (1996)
has  dispersion  $<$ 0.2 dex, the most 
remarkable case being [Si/Fe]  
with $\sigma$ = 0.11 dex in 12 DLA systems. 
The   compilation
of [Zn/Cr] data by Pettini et al. (1997) 
gives  $\sigma$ = 0.18 dex for 14 systems.
Within the typical measurement error
the observed abundances in the gas phase are essentially
constant. This  remarkable   result    
cannot be explained only by assuming that DLA systems
have equal intrinsic abundances  
and equal dust grains properties. Even in this case
the fractions in dust, and therefore also the observed abundances,
are expected to depend on the dust-to-metals ratio,
as can be  seen from Eq. (9). 
Since   metallicity and dust content can significantly
vary
among DLA systems, the only way to explain the constancy of
the observed abundances is that $k/Z \simeq$ is roughly
constant in different absorbers, as we indeed find.

\section{Discussion}

\subsection{The elemental abundance pattern of DLA absorbers}

Taken at face values, the positive
[Si/Fe] and [S/Fe] ratios and the negative  
[Mn/Fe] ratios observed
in DLA systems suggest an abundance pattern
similar to that found in the Galactic halo (Lu et al. 1996).
However, this interpretation requires that dust is completely
absent from the absorbers and   is unable to explain
why the iron-peak ratios [Zn/Fe] and [Ni/Fe] 
are different from zero.  
This inconsistency, together with the agreement between direct and
indirect estimates of the dust-to-gas ratios discussed in Section 4.3,
suggests that  any interpretation of the observed abundances must
take into account the presence of dust. 
 
In principle,  the dust in DLA absorbers could   differ 
from the dust in the Galaxy and an overall halo-like abundance
pattern could be hidden by the peculiar properties of the dust.  
Eq. (9) shows that one can
obtain a  halo-like   pattern from the observed quantities
by playing with the
chemical composition of the dust. Nevertheless,
since the condition $f_\X = (k/Z) (X/Y)_\intr^{-1} (\nu_X/c_{\rm e}) \leq 1$
must be satisfied, one will be also obliged to play with other parameters,
such as $c_{\rm e}$, in order to obtain meaningful results
(it is possible to see that this is in fact the case for Mn). 
Therefore, only by invoking a set of ad hoc requirements
on the chemical composition and on the extinction properties of the
dust grains
is it possible to argue that the intrinsic abundances
follow a halo-like pattern.   
By contrast, 
{\em the results shown in Fig. 1 have been derived  
without any fine tuning of the input parameters}, i.e.
the Galactic fractions in dust, 
{\em which are  known  from Galactic interstellar studies}. 
 
Our results indicate that all the DLA systems
for which good quality abundance determinations are available
show an elemental abundance pattern approximately solar,
within at most $\pm 0.2$ dex
(see however the comments on nitrogen in Section 4.2). 
The absorbers
included in this sample span a wide range
of metallicities ($ -1.4 \leq \log (Z/Z_{\sun}) \leq -0.1$)
and redshifts  ($  1 \leq z_{\rm abs}  \leq 2.5 $).
Even if Table 4 includes only 5 [Si/Fe] determinations,
the remarkably similar [Si/Fe] values reported
by Lu et al. (1996) in other systems without Zn
measurement, together with the approximately constant 
dust-to-metals ratio (Section 5) 
suggest that the lack of [$\alpha$/Fe] enhancement
is common to all the 12 absorbers with observed [Si/Fe] ratio.
Similarly, the 
mild [Mn/Fe] underabundance ($\simeq -0.15$ dex)
is probably present in all the 10 systems for which
measurements are available and not only in the 2 systems
shown in Table 4. 

Future measurements for a statistically significant number
of systems are required to see how general 
the solar-like ratios are. 
As mentioned in Section 4.2, nitrogen observations do show
deviations from solar relative abundances. 
In addition, DLA systems with intrinsically enhanced [$\alpha$/Fe] ratios
might be hidden by selection effects. 
For instance, damped absorbers with high  dust content
could systematically obscure the background QSO (Pei \& Fall 1995)
and this fact could affect the probability of detecting cases with 
enhanced [$\alpha$/Fe] ratios.

\subsection{DLA absorbers as dwarf galaxies}

Similarities between properties observed in DLA absorbers and in
gas rich dwarf galaxies, including the presence of large amounts of gas,
the low level of metallicity, 
and the complexity of the line profiles had already been pointed out
by York et al. (1986). 
Pettini, Boksenberg \& Hunstead (1990) noted, in addition,
a similarity in the star formation rate  and  
suggested that DLA absorbers might
arise predominantly in dwarf galaxies whose properties have not
changed significantly from $z=2.5$ to the present epoch.
According to Tyson (1988)
the frequency of detection of DLA absorbers can be naturally
explained by dwarf galaxies 
with no need to invoke extremely large HI diameters for the progenitors
of present-day spirals.   
Recent studies of the luminosity function of galaxies based on
deep field observations (Campos 1997; Zucca et al. 1997)
indicate a steepening of the faint end
of the luminosity function  
consistent with that suggested by Tyson (1988).

In addition to the above arguments  
we note here that
(i)  
the general lack of detection of molecules in DLA
absorbers (see Levshakov et al. 1992 and refs. therein)
is consistent with the low molecular content 
apparently typical of dwarf galaxies 
(Israel, Tacconi \& Baas 1995);
(ii) 
the lack of the extinction bump at $\simeq$ 2175 \AA\ in DLA absorbers
(Pei, Fall, \& Bechtold 1991) is characteristic of   
the extinction curves of the Magellanic Clouds.  
Finally, we note that the high fraction of metal-poor
stars predicted for the descendant of DLA absorbers
by  Lanzetta, Wolfe \& Turnshek  (1995) is consistent
with most DLA absorbers being  metal-poor dwarf galaxies.    
The high fraction of metal-poor stars predicted by the model
of  Lanzetta, Wolfe \& Turnshek  (1995) 
is not seen in the Milky Way
and this inconsistency is known as the "cosmic G-dwarf problem". 
To solve this problem it has been suggested
that a large fraction of the gas in DLA absorbers is consumed in forming 
stars in galactic bulges and spheroids rather than in galactic disks.
However, this explanation conflicts with the low mass fraction
of  the halo and of the bulge (Bahcall \& Soneira 1980; Rich 1992)
relative to the total mass
of the Milky Way, and is not supported by the range
of abundances recently found in the Galactic bulge  
(Rich 1992). 
If DLA absorbers are predominantly dwarf galaxies,
then there is no "cosmic G-dwarf problem" because  the stars
of the descendant of DLA absorbers do have low metallicities. 
We discuss now the observational evidences concerning
the nature of  DLA galaxies related to 
the results of the present work.

\subsubsection{Elemental abundance pattern }

The present results suggest  that DLA galaxies  
commonly show a nearly solar  [$\alpha$/Fe] ratio
at a low level of metallicity.
This property 
{\em does not show evolution} 
over a wide range of metallicities
($ -1.4 \leq \log (Z/Z_{\sun}) \leq -0.1$)
and redshifts  ($  1 \leq z_{\rm abs}  \leq 2.5 $). 
The lack of evolution up to high levels
of metallicity might suggest that the population of absorbers
is essentially unevolved to the present time.
This possibility, however, can only be investigated by studying a  
sample of low redshift DLA absorbers. 
 
The nearly solar  [$\alpha$/Fe] ratios  
do not support an origin 
{\em in spirals like the Milky Way, where the
enhancement of the }  [$\alpha$/Fe] 
{\em ratio is found not only in halo stars }
(Ryan et al. 1996),
{\em but also  in metal-poor stars of the disk }
(Edvardsson et al. 1993). 
It is unlikely therefore that DLA absorbers arise in massive
protogalactic disks.

Galaxies with low metallicities and without [$\alpha$/Fe] enhancement
are known to exist, as demonstrated  for instance
by abundance studies of the Magellanic Clouds  
(Wheeler, Sneden \& Truran 1989; p. 315).
Models of chemical evolution developed to explain  the elemental abundances
observed in blue compact and  dwarf irregular galaxies show that 
it is possible to obtain  low [$\alpha$/Fe] ratios at low metallicities
when many episodes of star formation and differential galactic winds
are assumed (Marconi, Matteucci \& Tosi 1994).
Similar models have   been applied to explain  the
abundances observed in some DLA systems (Matteucci, Molaro \& Vladilo 1997).

\subsubsection{Metallicity}

From a recent analysis Pettini et al. (1997b)  
have concluded that the metallicity distribution of DLA absorbers 
is different from that found in   
the stellar populations of the Milky Way. 
As we mentioned in Section 4.4, the dust-corrected metallicities
derived here are essentially equal to the direct
[Zn/H] measurements and therefore the conclusions
of Pettini et al. (1997b) are unchanged. 

\subsubsection{Dust content}

The dust-to-metals ratios of DLA absorbers
are systematically lower than in the Milky Way (Section 4.1), even at
metallicities comparable to that attained by the Galactic disk.
Similarly low dust-to-metals ratios  
are instead found in the Magellanic Clouds. 
This can be seen by combining the typical metallicities 
of the LMC and SMC,  
[Fe/H] = $-0.30$ and $-0.65$, 
respectively (Wheeler, Sneden \& Truran 1989; p. 315),
with their typical  gas-to-dust ratios,
$N_\HI$/E(B-V)  $ = 2 \times 10^{22}$ and  $5 \times 10^{22}$
cm$^{-2}$ mag$^{-1}$, respectively (Lequex 1989). 
These last ratios yield
$\ktilde = $ 0.24 and 0.096, respectively, once normalized
to the standard Galactic value $N_\HI$/E(B-V) $ = 4.8 \times 10^{21}$
cm$^{-2}$ mag$^{-1}$.  
The  Magellanic ratios follow remarkably well
the behaviour typical of DLA systems, as we show in Fig. 2,
and this is consistent again  with
an origin of DLA absorbers in dwarf galaxies. 

This result is at variance with the claim by
Roth \& Blades (1997)   that the Magellanic dust-to-metals ratios
are closer to   Milky Way  rather than  to DLA values. 
The claim is based on three measurements of the
Magellanic [Cr/Zn] ratio, two of which are in fact
consistent, within the error bars, with  DLA values. 
The third measurement, towards Sk $-68^\circ$73, 
gives a [Cr/Zn] ratio lower 
than in DLA systems, but in this case 
the Magellanic ZnII and CrII column densities  
are exceptionally high, not only compared to the other two
Magellanic lines of sight, but even compared to  
the 20 Galactic lines of sight observed  
with the HST GHRS (Roth \& Blades 1995). 
The LMC path toward Sk $-68^\circ$73 is clearly atypical
and might hide an exceptional amount of dust.
The few available Magellanic [Cr/Zn] measurements, therefore, do not  
contradict our conclusion that   Magellanic dust-to-metals ratios
 are  comparable to   DLA ratios.   

\subsubsection{Imaging of foreground galaxies}

A direct answer on the morphology of DLA galaxies can be obtained,
in principle, from imaging observations of the  close environment of  QSOs 
showing  DLA absorption.
A study of this kind has been performed by Le Brun et al. (1997)
for 8 absorbers at $z \leq 1$
in  7 fields observed with the HST/WFPC2. 
The candidate absorbing galaxies display a variety of
morphological types,  including 3 compact
galaxies, 2 low surface brightness  (LSB), and 3 spirals.
Our sample of Table 3 includes four 
of the absorbers studied by  Le Brun et al. (1997):
three are compact galaxies and one is a LSB galaxy.
The sample of dust-corrected
elemental abundances (Table 4) includes two of these absorbers and
both are compact galaxies.
Therefore, also the available imaging data are consistent with an
origin in dwarf galaxies of the absorbers of the present sample.

\subsubsection{Kinematics}

Also the kinematics gives important clues for understanding
the nature of DLA galaxies.
The velocity spread of the absorption components seen
through a randomly oriented galaxy along the line of sight
is in fact expected to be higher for a fastly rotating  
spiral than for a slowly rotating dwarf.  
Nevertheless, there are evidences of dwarfs
with velocity spreads higher than expected
from a slow rotation.
For instance, velocities spreads of 
$\approx$ 100 km s$^{-1}$ are   found both from stellar
(Westerlund 1990) and interstellar  
(Blades et al. 1988, Songaila et al. 1986) studies
of the Magellanic Clouds.
The compact galaxies 
identified by Le Brun et al. (1997)
as  the $z=0.859$ absorber toward QSO 0454+039
and the $z=1.766$ absorber toward QSO 1331+170
have velocity spreads of 150 km s$^{-1}$ 
and 100 km s$^{-1}$, respectively
(see Lu et al. 1996). 
Such relatively high velocity widths  
could be due to the presence of flows
or systematic motions 
superposed to the low rotation pattern  of dwarf galaxies
(see e.g. Sahu \& Blades 1997). 
Six out of the nine absorbers in our sample of Table 4
have velocity spreads of at most 100 - 150 km s$^{-1}$
and are therefore
consistent with an origin in   dwarf galaxies. 
The remaining three absorbers 
($z=2.462$ in QSO 0201+363,
$z=2.811$ in QSO 0528-250, and
$z=1.920$ in QSO 2206-199)
show instead 
velocities of 200 km s$^{-1}$ or even higher
(Lu et al. 1996, Prochaska \& Wolfe 1996, 
Prochaska \& Wolfe 1997). 
Such a high velocity width
is quite unlikely for  a single
dwarf galaxy, but   is not a conclusive evidence for a 
fastly rotating disk,  since multiple
intervening galaxies at similar redshift can also produce
a broad absorption profile.
In fact,  
Lu, Sargent \& Barlow (1997) present circumstantial evidence
that the large velocity width  
of the $z=2.811$ system toward QSO 0528-250 
can be due to the presence of more than one galaxy or sub-galactic
fragment along the line of sight. 
In our sample we do not see systematic differences in the elemental
abundance patterns between the cases with moderate 
and large velocity spread.
It is possible that the few cases with high velocity spread are
due to the superposition of low mass intervening galaxies.  
From a statistical analysis of sample of 17 DLA
absorbers  
Prochaska \& Wolfe (1997b) claim that models of slowly rotating
dwarf galaxies ($V_{\rm rot} = 50$ km s$^{-1}$)
are excluded at 97\% confidence level,
while models with rapidly rotating disks 
($V_{\rm rot} \simeq 250$ km s$^{-1}$)
are  consistent with the data at high confidence level. 
These authors do not exclude, however, 
that some of the cases with higher velocity spreads  
might arise from multiple intervening galaxies of low mass.  
An inspection of the velocity profiles of their  sample
reveals that the radial velocity spread is 
$\leq$ 100 km s$^{-1}$ in 10 systems, between
$\simeq$ 100 and $\simeq$ 150 km s$^{-1}$ in 2 systems, and 
$\geq$ 200  km s$^{-1}$ in only 5 systems.

\section{Conclusions}
 
In the present work
the general equations that link elemental
abundances observed in the gas phase with real abundances
of DLA absorbers (gas plus dust) have been presented.  
By assuming that dust is of Galactic type,  
simplified equations have been obtained 
in which the observed abundances are expressed as a function
of the dust-to-metals ratio, metallicity, 
element-to-element abundance pattern, and of
Galactic interstellar dust parameters, namely
the fractions in dust of the elements of interest.   

The   behaviour of the
dust depletions   in DLA absorbers 
has been investigated
for different sets of element-to-element abundances  
and of dust-to-metals ratios. 
The dust depletion pattern does not scale linearly
in logarithm with the Galactic pattern of depletion, contrary
to what assumed in some previous works. Instead, 
the differential depletion between pairs of elements tends to
vanish as the dust-to-metals ratio decreases. 
In addition, inversions in the differential depletion pattern
may appear if the DLA abundance pattern is not solar. 
 
A method has been presented for  determining
dust-to-metals ratios and dust-to-gas ratios in individual DLA absorbers 
from the abundances of elements with similar nucleosynthetic history.
This method has been applied to   
a sample of 17 DLA absorbers with available   Zn, Cr an Fe
measurements, assuming that these iron-peak elements  
trace each other down to the metallicities of
DLA systems, as they do in metal-poor stars of the Galaxy.
The resulting dust-to-gas ratios are mostly distributed
between 2\% and 25\% of the Galactic value, in good agreement
with the most probable range of dust-to-gas ratios
obtained by Pei, Fall \& Bechtold (1991)
from their reddening  study of QSOs with foreground DLA absorption. 

Dust-to-gas ratio and
metallicity are found to be correlated in DLA galaxies,
with a typical dust-to-metals ratio
$\simeq$ 60\% of the Galactic value.
This suggests that the metals injected from the stellar component
are systematically removed from   gas to   dust   
with an efficiency approximately constant but 
lower than in the Milky Way.  
The correlation between  dust-to-gas ratio and
metallicity  is relevant  to models of
cosmic chemical evolution   of the type investigated
by Pei \& Fall (1995).  
The approximate constancy of the dust-to-metals ratio 
explains why the element-to-element abundances observed in DLA
systems are remarkably constant in spite of the  
variable amount of dust in the absorbers. 

The resulting dust-to-metals ratios have been used
to correct from dust effects the element-to-element abundances
measured in a sub-sample of 9 DLA absorbers. 
The dust-corrected [$\alpha$/Fe] ratios do not show   
evidence of the   enhancement characteristic
of  metal-poor Galactic stars, but have instead
  solar values within $\pm 0.2$ dex. 
The [Mn/Fe] ratio shows a mild
underabundance, less marked than that  
found in metal-poor Galactic stars. The [Ni/Fe] ratio
is essentially solar, with a weak evidence for a modest
underabundance. These results rely on the assumption that
dust in DLA absorbers has same chemical composition as
Galactic interstellar dust.   
The available measurements of the [S/Zn] ratio, 
a proxy of the [$\alpha$/Fe] ratio virtually unaffected by dust, 
support the validity of the dust correction applied here 
(Molaro, Centuri\'on \& Vladilo, 1997).

The abundance pattern of DLA absorbers does not show
evidence of evolution 
in the range of redshifts  $1 \leq z_{\rm abs}  \leq 2.5$ 
and of metallicities $-1.4 \leq \log (Z/Z_{\sun}) \leq -0.1$
investigated. 
Studies of low redshift absorbers are required 
to establish whether evolution is present or not
closer to the present time (Vladilo et al. 1997a). 

The dust-corrected pattern of abundances indicates that
the chemical history of a large fraction of DLA galaxies is different from
that experienced by the Milky Way  halo or disk.
The results of the present work are consistent with
DLA absorbers being predominantly dwarf rather than spiral galaxies. 
Evidence in this sense comes from a wide spectrum of 
observational properties which include, in addition to the
abundance pattern, the dust-to-metals ratio and
the morphologycal type of the intervening galaxies
at $z \leq 1$. 
The kinematics of the absorbers is in most cases
consistent with an origin in dwarfs, even though
an origin in spirals gives a more natural
explanation of the highest observed velocity spreads. 

It is necessary to study a larger sample of DLA systems
by means of spectroscopic and imaging techniques
in order to determine with statistical accuracy  the fraction of spiral galaxies
in this class of QSO absorbers.  
At the same time, it is important to understand 
whether the low fraction of identified spirals is real or, instead,
is due to some selection effect or 
to a peculiar dust composition that would alter the
conclusions of the present work.  
If the fraction of spirals will be confirmed to be low, 
a mass of DLA galaxies lower than expected for proto-spirals objects 
would help to reconcile cold plus hot dark matter (CHDM) 
cosmology with observed properties of high redshift absorbers 
(see Klypin at al. 1995 and refs. therein).

\bigskip

The interpretation of this work has benefited from important
discussions with P. Molaro. I wish to thank
P. Bonifacio and M. Centuri\'on for their critical reading of the
manuscript. 
Discussions with S. Bardelli, S. Borgani, F. Matteucci, 
and S. Ortolani have helped to clarify specific topics of this
work. I also thank the referee, L. Lu, for his constructive remarks
and for anticipating some of his results in advance of publication.

\clearpage

\figcaption[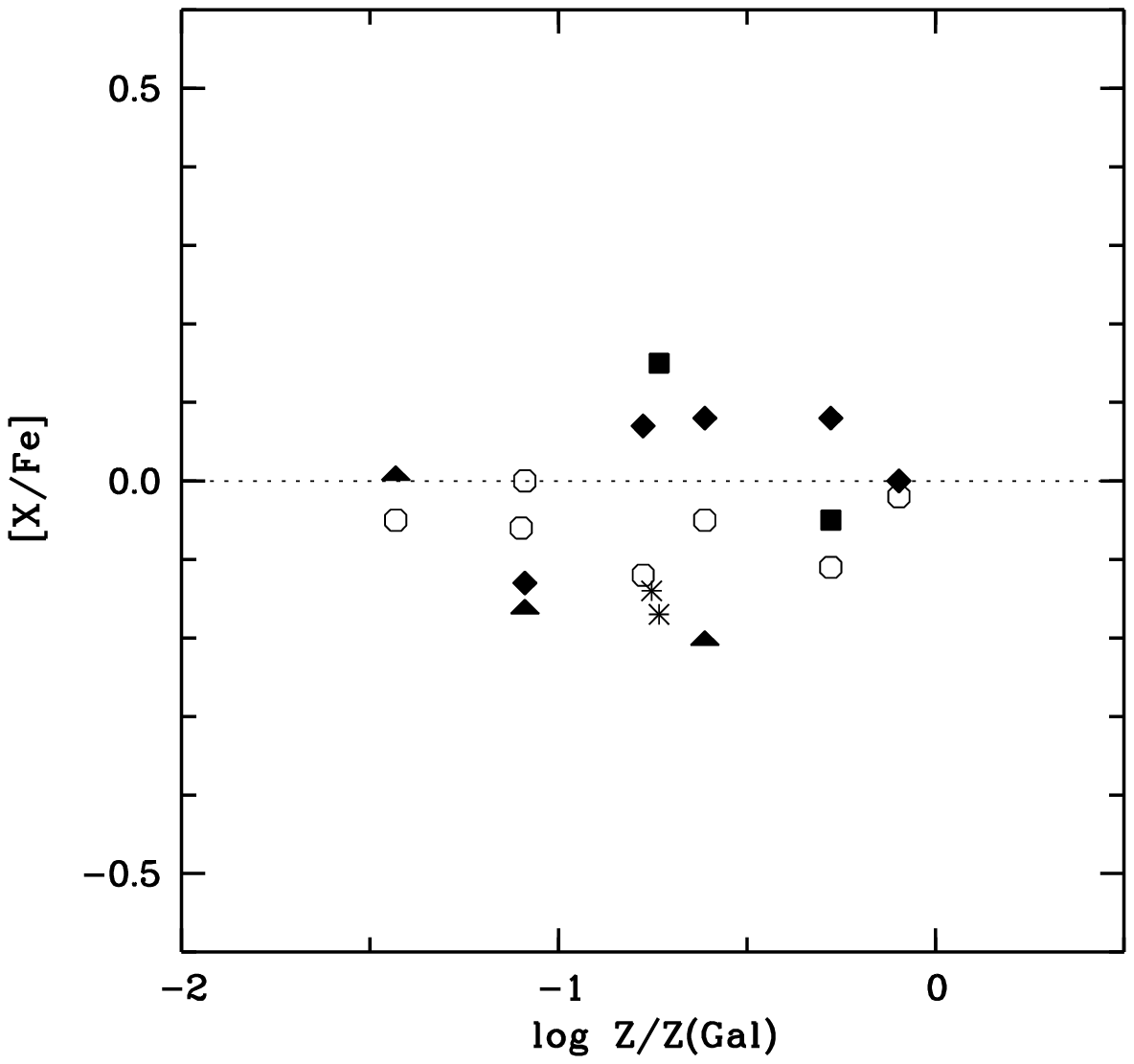]{
Dust-corrected relative abundances versus metallicity for the sample
of DLA absorbers shown in Table 4. Dotted line: solar reference abundance.
Filled diamonds, triangles and squares
indicate [Si/Fe], [S/Fe] and [Ti/Fe] ratios, respectively.
Asterisks:    [Mn/Fe] ratios.   
Open circles: [Ni/Fe] ratios.
Typical random errors of the ratios are of about $\pm 0.15$ dex. 
\label{fig1}}

\figcaption[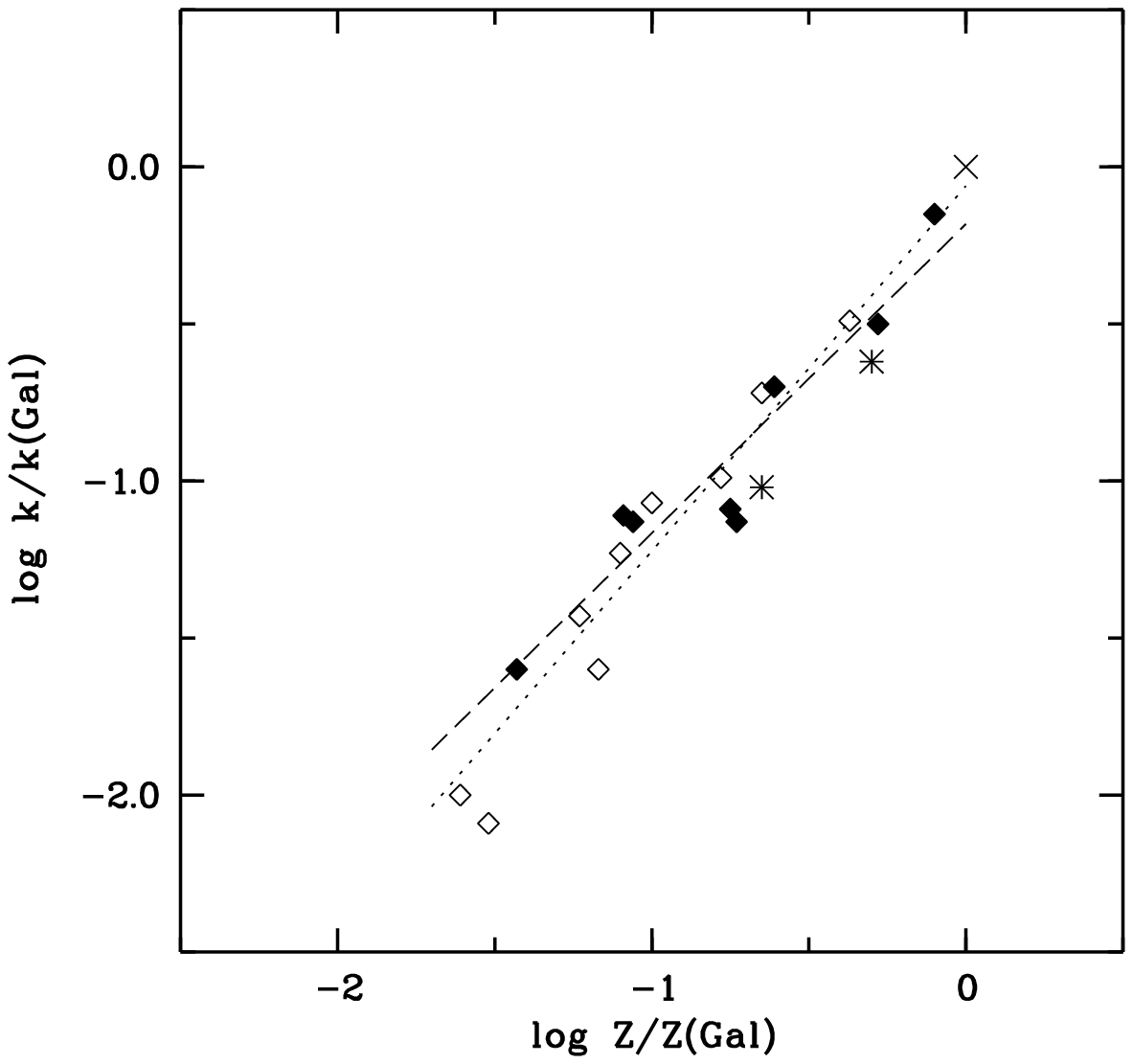]{
Dust-to-gas ratios versus metallicities normalized in Galactic units
for the sample of DLA absorbers shown in Table 3 (diamonds), 
the Magellanic Clouds
(asterisks) and the Galaxy (cross). Dashed line:  
linear regression through the full sample of 17 DLA absorbers.
Dotted line: linear regression through the 8 filled diamonds,
representing the DLA absorbers with
most precise measurements of the dust-to-gas ratio. 
 \label{fig2}}

\clearpage
 
\begin{deluxetable}{lcccccccc}
\footnotesize
\tablecaption{Predicted dust depletion patterns in Damped Ly$\alpha$ systems
              \label{tbl-1}  }
\tablewidth{0pt}
 
\tablehead{
    \colhead{ Element                                                 }                          
  & \colhead{ $f_{\X,\ism}$                          \tablenotemark{a} } 
  & \colhead{ [X/Fe]$_{\rm halo}$             \tablenotemark{b} }
  & \colhead{ S100                               \tablenotemark{c} }
  & \colhead{ S75                                 \tablenotemark{d} }
  & \colhead{ S50                                 \tablenotemark{e} }
  & \colhead{ H100                               \tablenotemark{f} }
  & \colhead{ H75                                 \tablenotemark{d} }
  & \colhead{ H50                                 \tablenotemark{e} }
 } 
  
\startdata

Si &  0.63 & +0.27 & -0.43 & -0.28 & -0.16 & -0.18 & -0.13 & -0.08 \nl
S  &  0.12 & +0.41 & -0.06 & -0.04 & -0.03 & -0.02 & -0.02 & -0.01 \nl
Ti &  0.91 & +0.22 & -1.02 & -0.49 & -0.26 & -0.35 & -0.23 & -0.14 \nl
Mn &  0.88 & -0.35 & -0.92 & -0.47 & -0.25 
&---\tablenotemark{g}&---\tablenotemark{g}& -1.82\nl
Cr &  0.92 & -0.07 & -1.10 & -0.51 & -0.27           
&---\tablenotemark{g}& -0.72 &  -0.34 \nl
Fe &  0.94 &  ---  & -1.22 & -0.53 & -0.28 & -1.22 & -0.53 & -0.28 \nl
Ni &  0.97 & -0.02 & -1.46 & -0.56 & -0.29
&---\tablenotemark{g}& -0.62 & -0.31 \nl
Zn &  0.35 & +0.04 & -0.19 & -0.13 & -0.08 & -0.17 & -0.12 & -0.08 \nl
 
\enddata

\tablenotetext{a}{Fractions in dust derived from Eq. (15) by adopting the   
                  dust depletions representative of Galactic warm disk
                  given by Savage \& Sembach (1996);
                  for Zn we adopt the average depletion for sight-lines with
                  lowest molecular content from Roth \& Blades (1995);
                  for Ti the base depletion from Jenkins (1987). }
 
\tablenotetext{b}{Average values in Galactic halo stars at a metallicity
                  level $Z/Z_{\sun}=0.1$; 
                  Si, Ti, Mn, Cr and Ni are estimated from the midmean
                  vector defined by Ryan et al. (1996);
                  S  from Francois (1988) corrected by --0.2 dex 
                  (Lambert 1989);
                  Zn from Sneden, Gratton \& Crocker (1991) }

\tablenotetext{c}{Predicted depletion pattern in DLA systems
                  $\delta_{\X,\dla} $ assuming
                  solar relative abundances,  
                  dust-to-metals ratio $\ktilde/\Ztilde=1.00$, 
                  and Galactic type dust with fractions in dust
                  as in column 2.} 

\tablenotetext{d}{Same as in previous column, except for 
                  dust-to-metals ratio $\ktilde/\Ztilde=0.75$. } 

\tablenotetext{e}{Same as in previous column, except for 
                  dust-to-metals ratio $\ktilde/\Ztilde=0.50$. } 

\tablenotetext{f}{Predicted depletion pattern in DLA systems
                  $\delta_{\X,\dla}$ assuming
                  halo-like   abundance as in column 3,
                  dust-to-metals ratio $\ktilde/\Ztilde=1.00$,
                  and Galactic type dust with fractions in dust
                  as in column 2;
                  iron depletion is the same as in the case of solar-like
                  abundances because iron is used as a reference for
                  the metallicity level.}
                  
\tablenotetext{g}{Solution not allowed.}

\end{deluxetable} 

\clearpage
 
\begin{deluxetable}{lccccccccccc}
\footnotesize
\tablecaption{Iron-peak abundance determinations in DLA systems. 
\tablenotemark{a,b} \label{tbl-2}}
\tablewidth{0pt}
\tablehead{
\colhead{QSO}     & \colhead{$z_{\rm abs}$} & 
\colhead{[Zn/H]}  & \colhead{ Ref.} &
\colhead{[Cr/H]}  & \colhead{ Ref.} &
\colhead{[Fe/H]}  & \colhead{ Ref.} &
\colhead{[Ni/H]}  & \colhead{ Ref.} & 
\colhead{[Mn/H]}  & \colhead{ Ref.}
} 
\startdata
0056 +014 & 2.777 & -1.23 & 1 & -1.35 & 1 & ---   &     &  ---  &     &   
  ---  &    \nl
0100 +130 & 2.309 & -1.55 & 2 & -1.79 & 2 & -1.95 & 3 & -2.13 & 2 & 
 ---   &   \nl
0112 +029 & 2.422 & -1.15 & 4,1 & -1.65 & 4,1 &  ---  &     &  ---  &     &     
  ---  &    \nl
0201 +363 & 2.462 & -0.26 & 5 & -0.90 & 5 & -0.80 & 5 & -1.00 &     &   
  ---  &    \nl
0302 -223 & 1.009 & -0.51 & 6 & -0.90 & 6 &  ---  &     &  ---  &     &     
  ---  &    \nl
0454 +039 & 0.859 & -0.83 & 7,1 & -1.02 & 7,1 & -0.97 & 8 &  ---  &     &     
-1.22  & 8  \nl
0458 -020 & 2.039 & -1.23 & 4,1 & -1.59 & 4,1 &  ---  &     & -1.90 & 9 &
  ---  &    \nl
0528 -250 & 2.811 & -0.76 & 8 & -1.23 & 8 & -1.23 & 8 & -1.56 & 8 &
  ---  &    \nl
0841 +129 & 2.374 & -1.35 & 1 & -1.64 & 1 &  ---  &     &  ---  &     &     
  ---  &    \nl
0935 +417 & 1.372 & -0.80 & 10,1  & -0.90 & 10,1 & -0.96 & 10 &  ---  &     &    
-1.24  & 10 \nl
1104 -180 & 1.661 & -0.80 & 11,1  & -1.29 & 11,1 &  ---  &     &  ---  &     &     
  ---  &    \nl
1151 +068 & 1.773 & -1.56 & 1 & -1.64 & 1 &  ---  &     &  ---  &     &     
  ---  &    \nl
1223 +175 & 2.465 & -1.68 & 4,1 & -1.82 & 4,1 &  ---  &     &  ---  &     &    
  ---  &    \nl
1328 +307 & 0.692 & -1.21 & 12,1& -1.66 & 12,1& -1.80 &     &  ---  &     &     
  ---  &    \nl
1331 +170 & 1.776 & -1.27 & 8 & -2.10 & 8 & -2.03 & 8 & -2.20 &     &     
  ---  &    \nl
2206 -199 & 1.920 & -0.38 & 13& -0.58 & 13& -0.68 & 13& -1.01 & 13&
  ---  &    \nl
2231 -001 & 2.066 & -0.88 & 8 & ---   &   &-1.14 & 8& -1.53 & 8 &
  ---  &    \nl
\enddata

\tablenotetext{a}
{All abundances have been normalized to the solar photospheric
values by Anders \& Grevesse (1989):
log\,(Zn/H)$_{\sun}$ = $-7.40$,
log\,(Cr/H)$_{\sun}$ = $-6.33$,
log\,(Ni/H)$_{\sun}$ = $-5.75$,
log\,(Mn/H)$_{\sun}$ = $-6.61$,
with the exception of
iron, taken from Hannaford et al. (1992):
log\,(Fe/H)$_{\sun}$ = $-4.52$ .
}
  
\tablenotetext{b} 
{Typical errors of these abundance determinations are in
the order of 0.1 dex;  
we refer to the below quoted bibliography for exact
  error determinations in individual cases.}

\tablenotetext{}{ 
REFERENCES: (1) Pettini et al. (1997); (2) Wolfe et al. (1994);
(3) Molaro, Centuri\'on, Vladilo (1997); (4) Pettini et al. (1994); 
(5) Prochaska \& Wolfe (1996); (6) Pettini et al. (1997b);
(7) Steidel et al. (1995); (8) Lu et al. (1996); 
(9) Meyer \& Roth (1990); (10) Meyer, Lanzetta \& Wolfe 1995;
(11) Smette et al. (1995); (12) Meyer \& York (1992);
(13) Prochaska \& Wolfe (1997); (14) Kulkarni et al. (1995)
}
 
\end{deluxetable}

\clearpage
 
\begin{deluxetable}{lccccccccc}
\footnotesize
\tablecaption{Dust-to-metal ratios, dust-to-gas ratios and metallicities
in Damped Lyman $\alpha$ systems
 \label{tbl-3}}
\tablewidth{0pt}
\tablehead{
\colhead{ QSO                                      }   & 
\colhead{ $z_{\rm abs}$                            }   & 
\colhead{ ${ \ktilde \over \Ztilde}_{\Zn,\Cr}$ 
          \tablenotemark{a}                        }   &
\colhead{ ${ \ktilde \over \Ztilde}_{\Zn,\Cr}$ 
          \tablenotemark{b}                        }   &  
\colhead{ ${ \ktilde \over \Ztilde}_{\Zn,\Fe}$ 
          \tablenotemark{a}                        }   & 
\colhead{ ${ \ktilde \over \Ztilde}_{\Zn,\Fe}$ 
          \tablenotemark{b}                        }   &                             
\colhead{ $\ktilde_{\Zn,\Cr}$    
          \tablenotemark{b}                        }   &                   
\colhead{ $\ktilde_{\Zn,\Fe}$   
          \tablenotemark{b}                        }   &  
\colhead{ $\log \ktilde$   
          \tablenotemark{b,c}                      }   &                            
\colhead{ $\log \Ztilde$    
          \tablenotemark{b,c}                      }                                                   
} 

\startdata
0056 +014 & 2.777 & 0.26 & 0.37 & ---  & ---  & 0.025 & ---   & -1.60 & -1.17  \nl
0100 +130 & 2.309 & 0.46 & 0.59 & 0.64 & 0.75 & 0.021 & 0.029 & -1.60 & -1.43  \nl
0112 +029 & 2.422 & 0.74 & 0.85 & ---  & ---  & 0.085 & ---   & -1.07 & -1.00  \nl
0201 +363 & 2.462 & 0.84 & 0.92 & 0.76 & 0.85 & 0.747 & 0.667 & -0.15 & -0.10  \nl
0302 -223 & 1.009 & 0.64 & 0.76 & ---  & ---  & 0.323 & ---   & -0.49 & -0.37  \nl
0454 +039 & 0.859 & 0.38 & 0.51 & 0.29 & 0.40 & 0.092 & 0.070 & -1.09 & -0.75  \nl
0458 -020 & 2.039 & 0.61 & 0.74 & ---  & ---  & 0.059 & ---   & -1.23 & -1.10  \nl
0528 -250 & 2.811 & 0.72 & 0.83 & 0.70 & 0.81 & 0.203 & 0.196 & -0.70 & -0.61  \nl
0841 +129 & 2.374 & 0.53 & 0.65 & ---  & ---  & 0.038 & ---   & -1.42 & -1.23  \nl
0935 +417 & 1.372 & 0.22 & 0.32 & 0.33 & 0.44 & 0.057 & 0.083 & -1.13 & -0.73  \nl
1104 -180 & 1.661 & 0.73 & 0.84 & ---  & ---  & 0.189 & ---   & -0.72 & -0.65  \nl
1151 +068 & 1.773 & 0.18 & 0.27 & ---  & ---  & 0.008 & ---   & -2.09 & -1.52  \nl
1223 +175 & 2.465 & 0.30 & 0.41 & ---  & ---  & 0.010 & ---   & -2.00 & -1.61  \nl
1328 +307 & 0.692 & 0.70 & 0.81 & 0.79 & 0.88 & 0.070 & 0.078 & -1.13 & -1.06  \nl
1331 +170 & 1.776 & 0.93 & 0.98 & 0.88 & 0.94 & 0.081 & 0.076 & -1.11 & -1.09  \nl
2206 -199 & 1.920 & 0.40 & 0.53 & 0.53 & 0.65 & 0.271 & 0.354 & -0.50 & -0.28  \nl
2231 -001 & 2.066 & ---  & ---  & 0.48 & 0.60 & ---   & 0.101 & -0.99 & -0.78  \nl
\enddata

\tablenotetext{a}{Values derived by assuming zinc completely
                  undepleted in Galactic interstellar gas,
                  i.e. $f_{\Zn,\ism}=0$      }
 
\tablenotetext{b}{Values derived by assuming zinc mildy
                  depleted in Galactic interstellar gas,
                  with $f_{\Zn,\ism}=0.35$   } 
                  
\tablenotetext{c}{Adopted value in computing the relation between dust-to-gas
                  ratio and metallicity      }  
 
\end{deluxetable}

\clearpage
 
\begin{deluxetable}{lccccccccccc}
\footnotesize
\tablecaption{Observed and dust-corrected relative abundances. 
\tablenotemark{a,b} \label{tbl-4}}
\tablewidth{0pt}
\tablehead{
\colhead{QSO}      & \colhead{$z_{\rm abs}$} & 
\colhead{[Si/Fe]}  & \colhead{ Ref.} &
\colhead{[S/Fe] }  & \colhead{ Ref.} &
\colhead{[Ti/Fe]}  & \colhead{ Ref.} &
\colhead{[Mn/Fe]}  & \colhead{ Ref.} & 
\colhead{[Ni/Fe]}  & \colhead{ Ref.}
} 
\startdata
 
0100 +130 & 2.309 &  --- &   & +0.41& 3 &  --- &   &  --- &   & -0.18&2,3\nl
          &       &  --- &   & +0.01&   &  --- &   &  --- &   & -0.05&   \nl
0201 +363 & 2.462 & +0.42& 5 & ---  &   &  --- &   &  --- &   & -0.20& 5 \nl
          &       &  0.00&   & ---  &   &  --- &   &  --- &   & -0.02&   \nl
0454 +039 & 0.859 &  --- &   & ---  &   &  --- &   & -0.25& 8 & ---  &   \nl
          &       &  --- &   & ---  &   &  --- &   & -0.14&   & ---  &   \nl
0458 -020 & 2.039 &  --- &   & ---  &   &  --- &   &  --- &  &(-0.67)\tablenotemark{c}&1,9\nl
          &       &  --- &   & ---  &   &  --- &   &  --- &   & -0.06&   \nl
0528 -250 & 2.811 & +0.48& 8 & +0.35& 8 &  --- &   &  --- &   & -0.33& 8 \nl
          &       & +0.08&   & -0.20&   &  --- &   &  --- &   & -0.05&   \nl
0935 +417 & 1.372 &  --- &   & ---  &   & +0.23& 10& -0.28& 10& ---  &   \nl
          &       &  --- &   & ---  &   & +0.15&   & -0.17&   & ---  &   \nl
1331 +170 & 1.776 & +0.18& 8 & +0.76&8,14&  --- &   &  --- &   & -0.17& 8 \nl
          &       & -0.13&   & -0.16&   &  --- &   &  --- &   &  0.00&   \nl
2206 -199 & 1.920 & +0.27& 13& ---  &   & -0.10& 13&  --- &   & -0.34& 13\nl
          &       & +0.08&   & ---  &   & -0.05&   &  --- &   & -0.11&   \nl
2231 -001 & 2.066 & +0.26& 8 & ---  &   &  --- &   &  --- &   & -0.39& 8 \nl
          &       & +0.07&   & ---  &   &  --- &   &  --- &   & -0.12&   \nl
\enddata

\tablenotetext{a}{For each damped absorber we give the observed, gas-phase
  abundance in the first row and the dust-corrected abundance in the second row;
  reference numbers are as in Table 2.}
  
\tablenotetext{b}  
{Abundances are relative  to the solar photospheric
values by Anders \& Grevesse (1989):
log\,(Si/H)$_{\sun}$ = $-4.45$,
log\,(S/H)$_{\sun}$  = $-4.79$,
log\,(Ti/H)$_{\sun}$ = $-7.01$,
log\,(Ni/H)$_{\sun}$ = $-5.75$,
log\,(Mn/H)$_{\sun}$ = $-6.61$,
with the exception of
iron, taken from Hannaford et al. (1992):
log\,(Fe/H)$_{\sun}$ = $-4.52$ .
}

\tablenotetext{c}{[Ni/Zn] used instead of [Ni/Fe] which is not available;
  the resulting [Ni/Zn] corrected for dust is, however, equal to [Ni/Fe]
  since the intrinsic [Zn/Fe] ratio is solar.}

\end{deluxetable}

\clearpage

\plotone{f1.eps}

\clearpage

\plotone{f2.eps}

\end{document}